\documentclass{PoS}

\title{Flavour anomalies in $B$ decays at LHCb}

\ShortTitle{$B$ Flavour Anomalies}

\author{\speaker{A.\ Hicheur}\thanks{On behalf of the LHCb collaboration - also at U.Constantine.}\\
        Federal University of Rio de Janeiro.\\
        E-mail: \email{hicheur@if.ufrj.br}}

\abstract{The direct searches for Beyond Standard Model (BSM) particles have been constraining their mass scale to the extent where it is now becoming consensual that such particles are likely to be above the energy reach of the LHC. Meanwhile, the studies of indirect probes of BSM physics, with all their diversity, have been progressing both in accurracy and in setting up observables with reduced theoretical uncertainties. The observation of flavour anomalies in $b$ hadron decays represents an important part of the program of indirect detection of BSM physics. Several benchmark analyses involving leptonic or semileptonic decays are presented, with an emphasis on intriguing patterns which are systematic in their trend, though not individually significant yet.}

\FullConference{The 21st international workshop on neutrinos from accelerators (NuFact2019)\\
		August 26 - August 31, 2019\\
		Daegu, Korea}

\usepackage{ifthen} 
\newboolean{pdflatex}
\setboolean{pdflatex}{true} 

\newboolean{articletitles}
\setboolean{articletitles}{true} 

\newboolean{uprightparticles}
\setboolean{uprightparticles}{false} 

\newboolean{inbibliography}
\setboolean{inbibliography}{false} 

\usepackage{microtype}
\usepackage{lineno}  
\usepackage{xspace} 
\usepackage{caption} 

\usepackage{graphicx,subfigure}  
\usepackage{color}
\usepackage{colortbl}
\graphicspath{{./figs/}} 

\usepackage{amsmath} 
\usepackage{amssymb}
\usepackage{amsfonts}
\usepackage{upgreek} 

\newcommand*\patchAmsMathEnvironmentForLineno[1]{%
\expandafter\let\csname old#1\expandafter\endcsname\csname #1\endcsname
\expandafter\let\csname oldend#1\expandafter\endcsname\csname
end#1\endcsname
 \renewenvironment{#1}%
   {\linenomath\csname old#1\endcsname}%
   {\csname oldend#1\endcsname\endlinenomath}%
}
\newcommand*\patchBothAmsMathEnvironmentsForLineno[1]{%
  \patchAmsMathEnvironmentForLineno{#1}%
  \patchAmsMathEnvironmentForLineno{#1*}%
}
\AtBeginDocument{%
\patchBothAmsMathEnvironmentsForLineno{equation}%
\patchBothAmsMathEnvironmentsForLineno{align}%
\patchBothAmsMathEnvironmentsForLineno{flalign}%
\patchBothAmsMathEnvironmentsForLineno{alignat}%
\patchBothAmsMathEnvironmentsForLineno{gather}%
\patchBothAmsMathEnvironmentsForLineno{multline}%
\patchBothAmsMathEnvironmentsForLineno{eqnarray}%
}

\usepackage{url}    

\usepackage{xspace} 
\usepackage{upgreek}

\def\MagUp {\mbox{\em Mag\kern -0.05em Up}\xspace}

\ifthenelse{\boolean{uprightparticles}}%
{

 \def\Pmu         {\ensuremath{\upmu}\xspace}

 \def\Ppi         {\ensuremath{\uppi}\xspace}

 \def\Ppsi        {\ensuremath{\uppsi}\xspace}

 \def\PDelta      {\ensuremath{\Delta}\xspace}                 
 \def\PXi      {\ensuremath{\Xi}\xspace}                 
 \def\PLambda      {\ensuremath{\Lambda}\xspace}                 
 \def\PSigma      {\ensuremath{\Sigma}\xspace}                 
 \def\POmega      {\ensuremath{\Omega}\xspace}                 
 \def\PUpsilon      {\ensuremath{\Upsilon}\xspace}                 
 
 \def\PB      {\ensuremath{\mathrm{B}}\xspace}                 
                  
 \def\PD      {\ensuremath{\mathrm{D}}\xspace}

 \def\PJ      {\ensuremath{\mathrm{J}}\xspace}                 
 \def\PK      {\ensuremath{\mathrm{K}}\xspace}

 \def\Pb      {\ensuremath{\mathrm{b}}\xspace}                 
 \def\Pc      {\ensuremath{\mathrm{c}}\xspace}

 \def\Pi      {\ensuremath{\mathrm{i}}\xspace}

 \def\Ps      {\ensuremath{\mathrm{s}}\xspace}

}
{

 \def\Pmu         {\ensuremath{\mu}\xspace}

 \def\Ppi         {\ensuremath{\pi}\xspace}

 \def\Ppsi        {\ensuremath{\psi}\xspace}                 
                  
 \mathchardef\PDelta="7101
 \mathchardef\PXi="7104
 \mathchardef\PLambda="7103
 \mathchardef\PSigma="7106
 \mathchardef\POmega="710A
 \mathchardef\PUpsilon="7107
                  
 \def\PB      {\ensuremath{B}\xspace}                 
                  
 \def\PD      {\ensuremath{D}\xspace}

 \def\PJ      {\ensuremath{J}\xspace}                 
 \def\PK      {\ensuremath{K}\xspace}

 \def\Pb      {\ensuremath{b}\xspace}                 
 \def\Pc      {\ensuremath{c}\xspace}

 \def\Pi      {\ensuremath{i}\xspace}

 \def\Ps      {\ensuremath{s}\xspace}

}

\makeatletter
\ifcase \@ptsize \relax
  \newcommand{\miniscule}{\@setfontsize\miniscule{4}{5}}
\or
  \newcommand{\miniscule}{\@setfontsize\miniscule{5}{6}}
\or
  \newcommand{\miniscule}{\@setfontsize\miniscule{5}{6}}
\fi
\makeatother

\DeclareRobustCommand{\optbar}[1]{\shortstack{{\miniscule (\rule[.5ex]{1.25em}{.18mm})}
  \\ [-.7ex] $#1$}}

\def\mup        {{\ensuremath{\Pmu^+}}\xspace}
\def\mun        {{\ensuremath{\Pmu^-}}\xspace} 

\def\squark    {{\ensuremath{\Ps}}\xspace}

\def\cquark    {{\ensuremath{\Pc}}\xspace}

\def\bquark    {{\ensuremath{\Pb}}\xspace}

\def\pion   {{\ensuremath{\Ppi}}\xspace}

\def\pip    {{\ensuremath{\pion^+}}\xspace}

\def\kaon    {{\ensuremath{\PK}}\xspace}
  \def\Kbar    {{\kern 0.2em\overline{\kern -0.2em \PK}{}}\xspace}

\def\KorKbar    {\kern 0.18em\optbar{\kern -0.18em K}{}\xspace}

\def\Kp      {{\ensuremath{\kaon^+}}\xspace}
\def\Km      {{\ensuremath{\kaon^-}}\xspace}

  \def\Dbar    {{\kern 0.2em\overline{\kern -0.2em \PD}{}}\xspace}

\def\DorDbar    {\kern 0.18em\optbar{\kern -0.18em D}{}\xspace}

\def\B       {{\ensuremath{\PB}}\xspace}
\def\Bbar    {{\ensuremath{\kern 0.18em\overline{\kern -0.18em \PB}{}}}\xspace}

\def\BorBbar    {\kern 0.18em\optbar{\kern -0.18em B}{}\xspace}

\def\Bu      {{\ensuremath{\B^+}}\xspace}

\def\Bp      {{\ensuremath{\Bu}}\xspace}

\def\Bs      {{\ensuremath{\B^0_\squark}}\xspace}

\def\Bc      {{\ensuremath{\B_\cquark^+}}\xspace}

\def\jpsi     {{\ensuremath{{\PJ\mskip -3mu/\mskip -2mu\Ppsi\mskip 2mu}}}\xspace}
\def\psitwos  {{\ensuremath{\Ppsi{(2S)}}}\xspace}

  \def\Y#1S{\ensuremath{\PUpsilon{(#1S)}}\xspace}

\def\Lz          {{\ensuremath{\PLambda}}\xspace}
\def\Lbar        {{\ensuremath{\kern 0.1em\overline{\kern -0.1em\PLambda}}}\xspace}
\def\LorLbar    {\kern 0.18em\optbar{\kern -0.18em \PLambda}{}\xspace}

\def\Lb      {{\ensuremath{\Lz^0_\bquark}}\xspace}

\def\to                 {\ensuremath{\rightarrow}\xspace}

\def\AT#1     {\ensuremath{A_{\mathrm{T}}^{#1}}\xspace}           

\def\C#1      {\ensuremath{\mathcal{C}_{#1}}\xspace}                       
\def\Cp#1     {\ensuremath{\mathcal{C}_{#1}^{'}}\xspace}                    
\def\Ceff#1   {\ensuremath{\mathcal{C}_{#1}^{\mathrm{(eff)}}}\xspace}        
\def\Cpeff#1  {\ensuremath{\mathcal{C}_{#1}^{'\mathrm{(eff)}}}\xspace}       
\def\Ope#1    {\ensuremath{\mathcal{O}_{#1}}\xspace}                       
\def\Opep#1   {\ensuremath{\mathcal{O}_{#1}^{'}}\xspace}                    

\newcommand{\tev}{\ifthenelse{\boolean{inbibliography}}{\ensuremath{~T\kern -0.05em eV}}{\ensuremath{\mathrm{\,Te\kern -0.1em V}}}\xspace}
\newcommand{\gev}{\ensuremath{\mathrm{\,Ge\kern -0.1em V}}\xspace}
\newcommand{\mev}{\ensuremath{\mathrm{\,Me\kern -0.1em V}}\xspace}
\newcommand{\kev}{\ensuremath{\mathrm{\,ke\kern -0.1em V}}\xspace}
\newcommand{\ev}{\ensuremath{\mathrm{\,e\kern -0.1em V}}\xspace}
\newcommand{\gevc}{\ensuremath{{\mathrm{\,Ge\kern -0.1em V\!/}c}}\xspace}
\newcommand{\mevc}{\ensuremath{{\mathrm{\,Me\kern -0.1em V\!/}c}}\xspace}
\newcommand{\gevcc}{\ensuremath{{\mathrm{\,Ge\kern -0.1em V\!/}c^2}}\xspace}
\newcommand{\gevgevcccc}{\ensuremath{{\mathrm{\,Ge\kern -0.1em V^2\!/}c^4}}\xspace}
\newcommand{\mevcc}{\ensuremath{{\mathrm{\,Me\kern -0.1em V\!/}c^2}}\xspace}

\def\gsim{{~\raise.15em\hbox{$>$}\kern-.85em
          \lower.35em\hbox{$\sim$}~}\xspace}
\def\lsim{{~\raise.15em\hbox{$<$}\kern-.85em
          \lower.35em\hbox{$\sim$}~}\xspace}

\def\tell1  {TELL1\xspace}
\def\ukl1   {UKL1\xspace}

\usepackage{cite} 
\usepackage{mciteplus}

\begin{document}
\section{Introduction}
Heavy Flavour decays are usually described by low energy effective Hamiltonians forming an Effective Field Theory (EFT) (see, e.g. reference \cite{Buchalla} for a review). The Hamiltonians are written as:
\begin{equation}
H = \sum_{i} V_{CKM}^i C_i(\mu) O_i(\mu),
\end{equation}
where $C_i(\mu)$ are the Wilson coefficients integrating out the physics above the scale $\mu$ (short range), $O_i(\mu)$ are current operators which matrix elements represent the low energy (non-perturbative/long range) hadronic physics, and $\mu$ is the renormalization scale (typically $\sim 1 \gev$) distinguishing the two regimes. $V_{CKM}^i$ represents the flavour coupling associated to an operator $O_i$, i.e., Cabibbo-Kobayashi-Maskawa (CKM) matrix elements for SM operators.\\
For the semileptonic tree decays, contrary to SM where the coupling of the mediating $W$ boson does not discriminate between lepton flavours, a BSM mediator might exhibit different couplings between light and heavy leptons. This is Lepton Flavour Universality Violation (LFUV). Such an effect could also occur for the semileptonic loop decays $b\to s\ell\ell$ where the dominant operators are $O_7,~O_9,~O_{10}$. Loop decays could also be the ground of new dynamics involving Lepton Flavour Violation (LFV) where leptons of different flavours are produced together.
\section{Semileptonic tree decays}
\begin{figure}[htb]
\begin{center}
\includegraphics[width=0.4\textwidth]{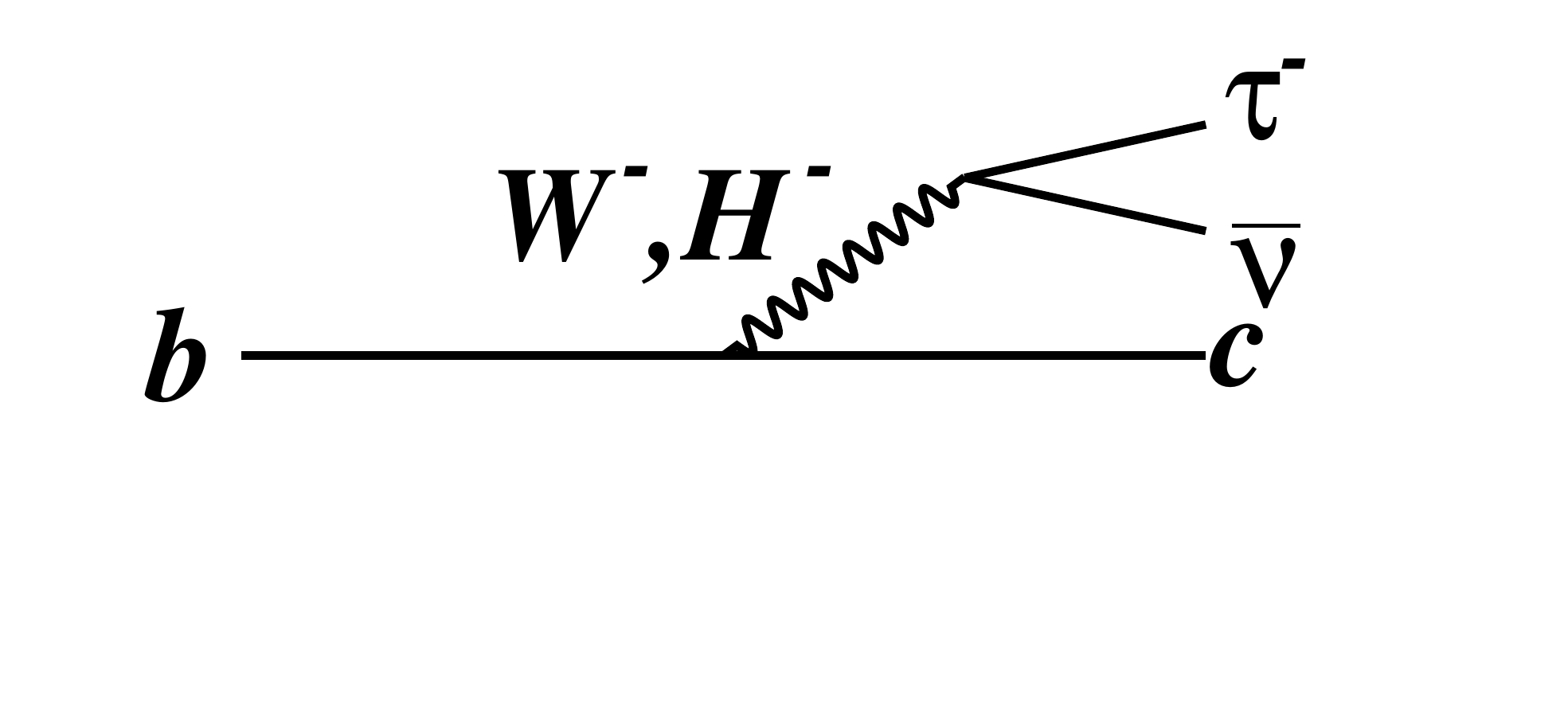}
\end{center}
\vspace{-1.3cm}
\caption{$b\to c\tau^-\bar\nu$ semileptonic transition with a mediating $W$ or charged Higgs boson.}
\label{Fig:btoctau_Higgs}
\end{figure}

Such decays are generically written as $H_b\to H_c \ell^- \bar\nu$ where $H_b$ is a $b$ hadron and $H_c$ is a charm hadron. The search for a possible LFUV is sought through the measurement of the ratio:
\begin{equation}
R(H_c) = \frac{\mathcal{B}(H_b\to H_c \tau^- \bar{\nu})}{\mathcal{B}(H_b\to H_c \mu^- \bar{\nu})},
\end{equation}
where $\mathcal{B}$ denotes the branching fraction. A BSM mediating heavy boson might couple preferentially to the tau lepton, as in Fig.\ref{Fig:btoctau_Higgs}, and thus produce a $R(H_c)$ ratio different from the expected SM-based calculations.

\begin{figure}[htb]
\begin{center}
\includegraphics[width=0.59\textwidth]{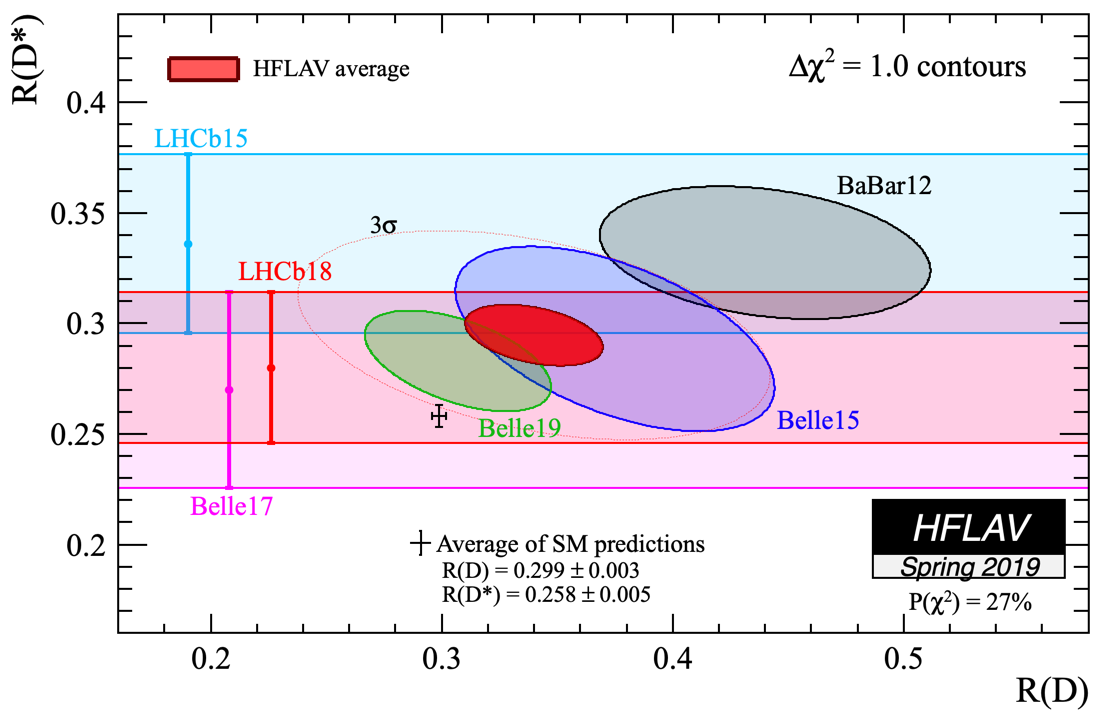}
\end{center}
\vspace{-0.7cm}
\caption{Average of $R(D)$ and $R(D^*)$.}
\label{Fig:rd_rdstar_hflav}
\end{figure}

The modes $\bar{B}^0 \to D^{+(*)} \ell^- \bar\nu$ have drawn a particular attention both at the $b$ factories and the LHCb experiment. The LHCb collaboration has been focusing so far on $R(D^{+*})$, through the decay chain $D^{*+}\to D^0(\to \Km\pip)\pip$. The $\tau$ lepton is reconstructed in the muonic mode, $\tau^-\to\mu^-\bar{\nu}_\mu\nu_\tau$ \cite{Dstar_MuMode}, or the hadronic mode $\tau\to\pi\pi\pi(\pi^0)\nu_\tau$ \cite{Dstar_HadMode}. The discriminating variables include the missing mass, $m_{miss}^2 = (P_B-P_{D^*}-P_\mu)^2$, the momentum transfer $q^2 = (P_B-P_{D^*})^2$, the muon energy $E_\mu^*$ and the $\tau$ lifetime (for the hadronic mode). 
The muonic tau analysis \cite{Dstar_MuMode} obtained a measurement of $R(D^*)=0.336\pm0.027(stat)\pm0.030(syst)$ while the hadronic tau study \cite{Dstar_HadMode} gives $R(D^*)=0.291\pm0.021(stat)\pm0.026(syst)\pm0.013(BF)$, where the last uncertainty is due to the external branching fraction of the normalizing channel. The latest HFLAV averaging \cite{hflav} in the $R(D)-R(D^*)$ plane, including the recent Belle collaboration $R(D^{(*)})$ measurements \cite{belle_rd_2019}, is shown in Fig.\ref{Fig:rd_rdstar_hflav}. Compared to an averaged series of SM-based predictions \cite{rd_theory}, a discrepancy of 3.1$\sigma$ is observed.

For the \Bc meson, a first $R(\jpsi)$ measurement has been performed recently by the LHCb collaboration for the $\tau$ muonic mode \cite{RJpsi}, leading to $R(\jpsi)=0.71\pm0.17(stat)\pm0.18(syst)$ which lies 2$\sigma$ above the range of the known theoretical estimates \cite{RJpsi_theory}.

\section{$b\to s\ell\ell$ transitions}
\begin{figure}[htb]
\begin{center}
\includegraphics[width=0.3\textwidth]{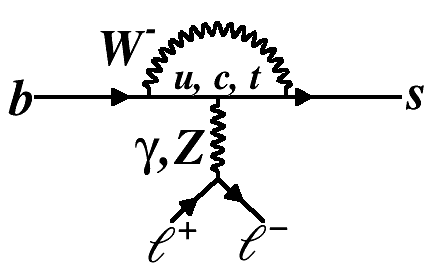}
\includegraphics[width=0.4\textwidth]{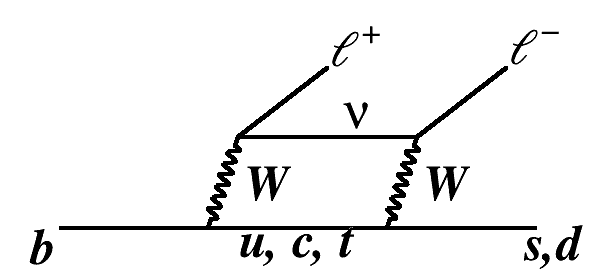}
\end{center}
\vspace{-0.5cm}
\caption{$b\to s\ell\ell$ (left) penguin and (right) box transitions.}
\label{Fig:btosll_diags}
\end{figure}

\begin{figure}[htb]
\begin{center}
\includegraphics[width=0.4\textwidth]{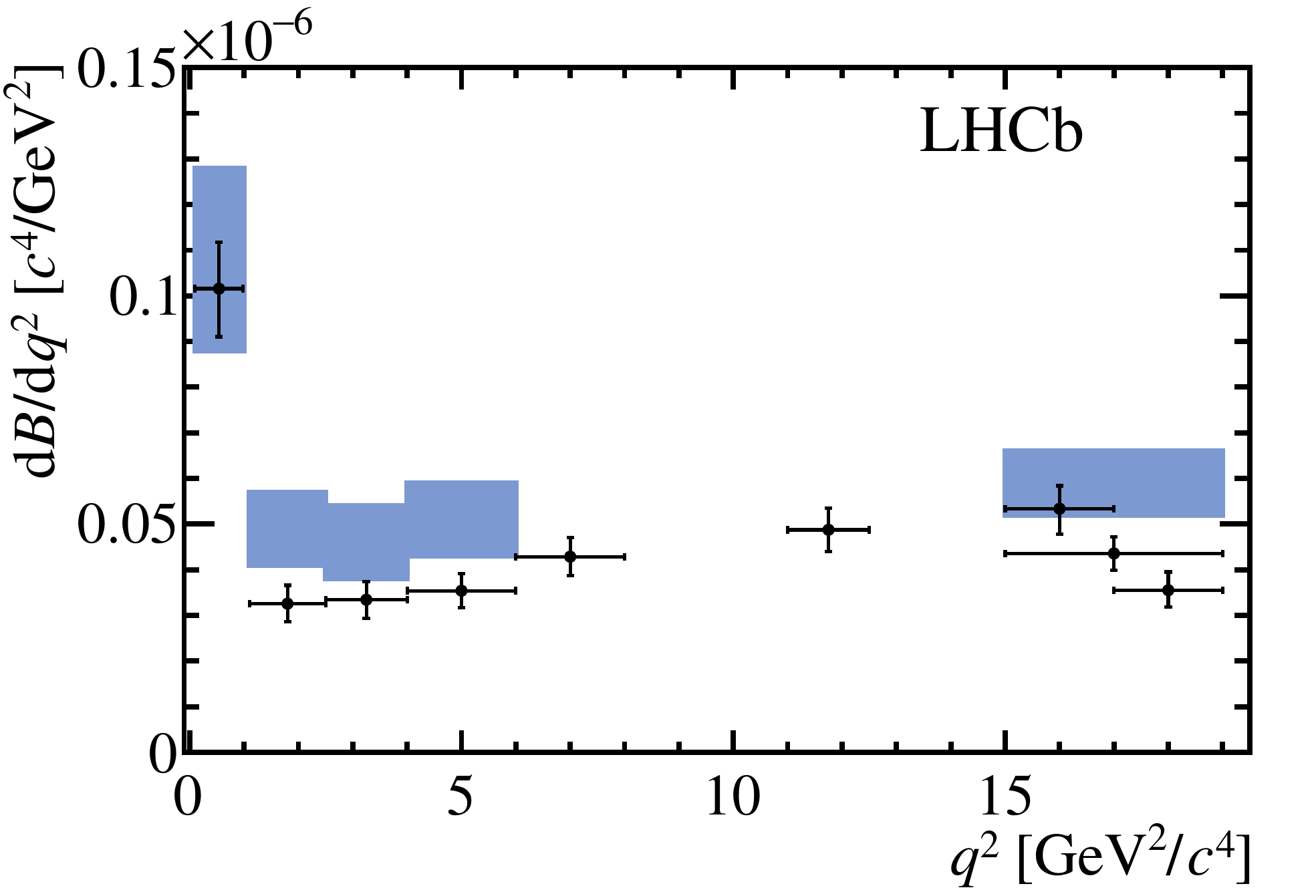}
\includegraphics[width=0.4\textwidth]{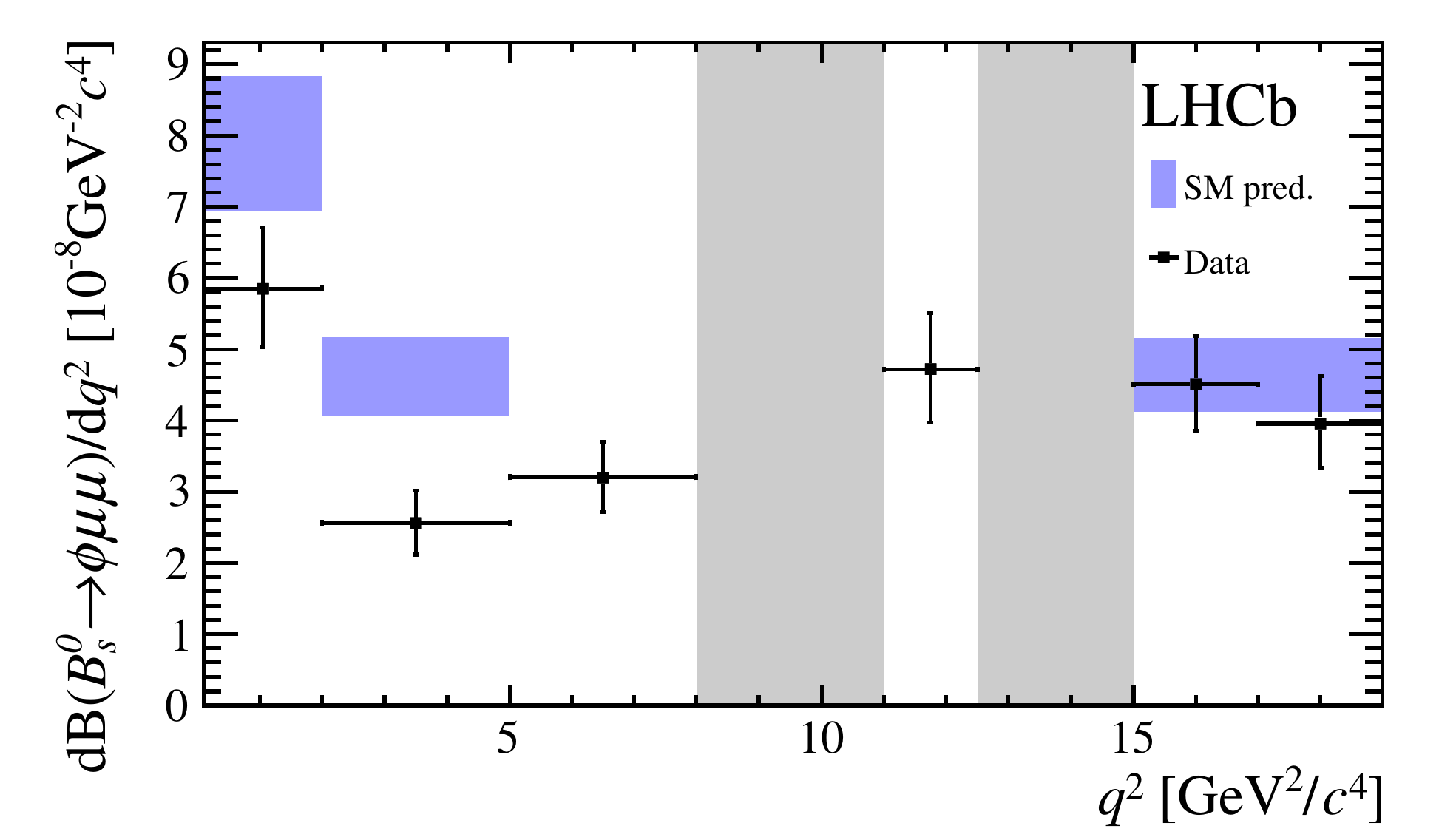}
\end{center}
\vspace{-0.5cm}
\caption{$\frac{d\Gamma}{dq^2}$ distribution for (left) $B^0\to K^{*0}\mup\mun$ and (right) $\Bs\to \phi\mup\mun$. The \jpsi and \psitwos bands are vetoed.}
\label{Fig:dGdq2}
\end{figure}

\begin{figure}[htb]
\begin{center}
\includegraphics[width=0.4\textwidth]{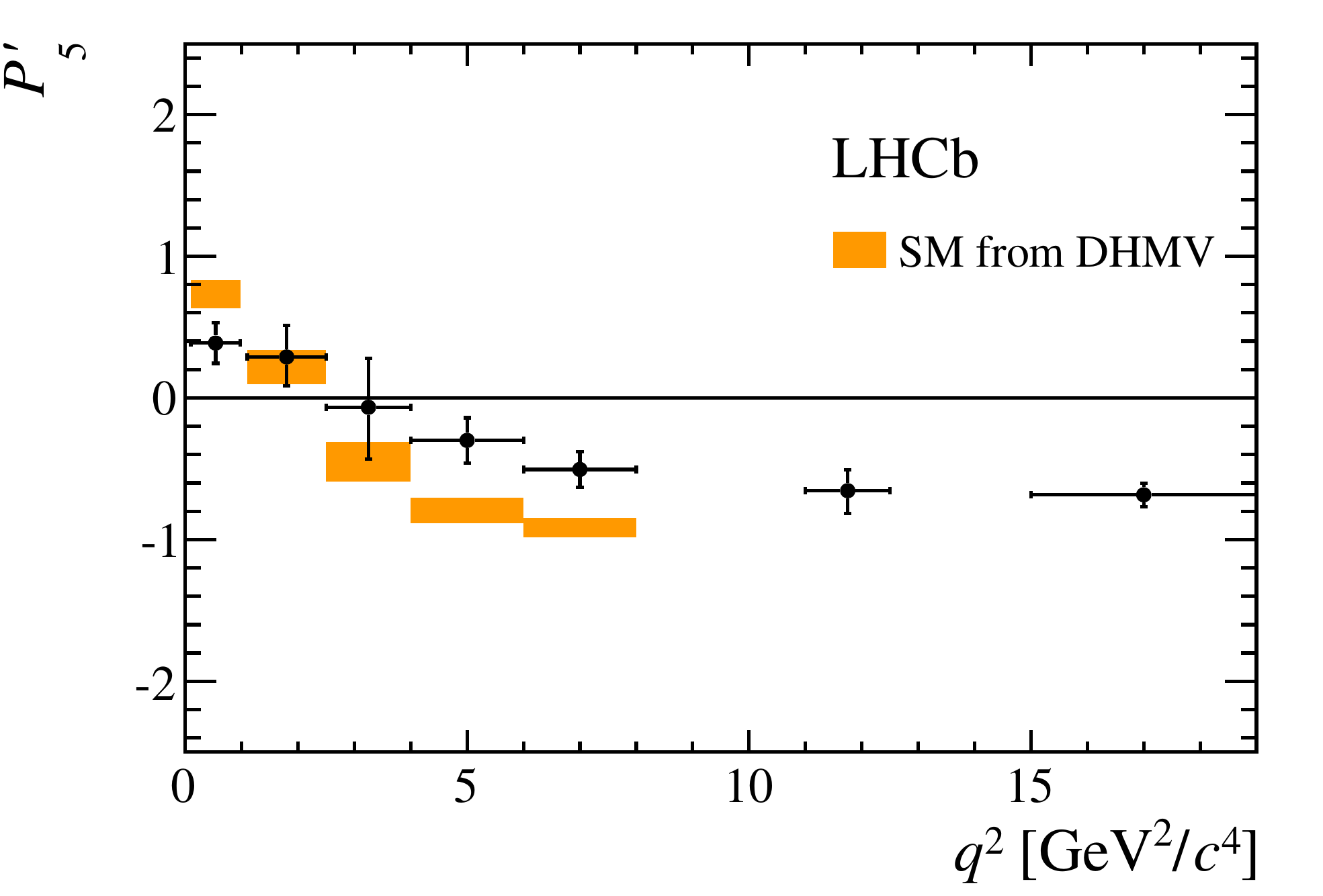}
\end{center}
\vspace{-0.5cm}
\caption{Evolution of $P_5^\prime$ (see text) as a function of $q^2$ for $B^0\to K^{*0}\mup\mun$.}
\label{Fig:P5Prime}
\end{figure}

These transitions proceed through the diagrams shown in Fig.\ref{Fig:btosll_diags} at quark level. The operators contributing to these decays are not evenly distributed in the $q^2 = m_{\ell\ell}^2$ range: at low $q^2$, $O_7$ dominates (for transitions to non-scalar hadrons), in the central $q^2$ region below the charmonium resonances, $O_7$ and $O_9$ interfere, and at high $q^2$ $O_9$ and $O_{10}$ interfere.
At the hadron level, the modes investigated by LHCb are $\Bp\to\Kp\ell^+\ell^-$, $B^0\to K^{0}\ell^+\ell^-$, $B^0\to K^{*0}\ell^+\ell^-$, $\Bs\to \phi\ell^+\ell^-$ and $\Lb\to \Lz\ell^+\ell^-$. 
A series of studies \cite{lhcb_xmumu} have dealt with the dynamics of the muonic modes, $\ell=\mu$, to infer the differential decay rate $\frac{d\Gamma}{dq^2}$, as illustrated in Fig.\ref{Fig:dGdq2}. The data is systematically below the SM-based theoretical predictions, with local discrepancies exceeding 3$\sigma$. Attempting to dig more into this intriguing behaviour, angular analyses were performed for $B^0\to K^{*0}\mup\mun$ \cite{lhcb_xmumu}(c), $\Bs\to \phi\mup\mun$ \cite{lhcb_xmumu}(b) and $\Lb\to \Lz\mup\mun$ \cite{lhcb_lzmumu_ang}. For $B^0\to K^{*0}\mup\mun$, the discrepancy reported in previous studies for the quantity $P_5^\prime = \frac{S_5}{\sqrt{F_L(1-F_L)}}$, built to reduce the hadronic uncertainties \cite{descotes} from the coefficients $S_5$ and $F_L$ (fraction of the $K^*$ longitudinal polarization) of the angular distribution, seems to be persistent as shown in Fig.\ref{Fig:P5Prime}.

Another way to probe the presence of New Physics is to measure the ratio $R_{H_s} = \frac{\mathcal{B}(H_b\to H_s\mup\mun)}{\mathcal{B}(H_b\to H_s e^+e^-)}$, where $H_s$ denotes a strange hadron. The LHCb collaboration studied $R_K$ \cite{lhcb_rk_ratios}(a) ($\Bp\to\Kp\ell^+\ell^-$) and $R_{K^*}$ \cite{lhcb_rk_ratios}(b) ($B^0\to K^{*0}\ell^+\ell^-$). For $R_K$, the explored $q^2$ range is $[1.1,6]~\gevgevcccc$, i.e. below the charmonium radiative tails, and above backgrounds of the type $\Bp\to\Kp\phi(\to\ell^+\ell^-)$. The $R_{K^*}$ analysis uses two bins in $q^2$, $[0.045,1.1]~\gevgevcccc$ (above the photon pole) and $[1.1,6]~\gevgevcccc$. 

The obtained measurements are $R_K=0.846^{+0.054}_{-0.060}(stat)^{+0.014}_{-0.016}(syst)$ ($1.1<q^2<6\gevgevcccc$); $R_{K^*}=0.66^{+0.11}_{-0.17}(stat)\pm0.03(syst)$ for $0.045<q^2<1.1\gevgevcccc$ and $0.69^{+0.11}_{-0.17}(stat)\pm0.05(syst)$ for $1.1<q^2<6\gevgevcccc$. All these values are systematically below the SM-based predictions by 2.2$\sigma$ to 2.5$\sigma$ \cite{RK_theo}.\\

The hints of LFUV in $b \to s\ell\ell$ decays have motivated a recent LFV search, $\Bp\to\Kp\mu^{\pm}e^{\mp}$ \cite{lhcb_kmue}, leading to the establishment of the 95\% confidence level (CL) limits: $\mathcal{B}(\Bp\to\Kp\mu^{-}e^{+})< 9.5\times10^{-9}$ and $\mathcal{B}(\Bp\to\Kp\mu^{+}e^{-})< 8.8\times10^{-9}$.

\section{$B\to\ell\ell$}
For these pure leptonic modes, the combination \cite{straub} of the most recent results of the ATLAS, CMS and LHCb experiments shows that $\Bs\to\mup\mun$ is 2$\sigma$ below the SM-based predictions. This does not include the very latest CMS result \cite{cms_bmumu}. On the LFV front, LHCb has recently published results on the search for the decays $B^0_{(s)}\to \tau^{\pm}\mu^{\mp}$ \cite{lhcb_taumu}, with the 95\% CL limits being $\mathcal{B}(B^0_{s}\to \tau^{\pm}\mu^{\mp})< 4.2\times10^{-5}$ and $\mathcal{B}(B^0\to \tau^{\pm}\mu^{\mp})< 1.4\times10^{-5}$.
\section{Summary}
The current state-of-the-art of the anomalies in the b hadron decays has not converged yet to a individual measurement reaching the observation level of an involvement of BSM physics. The combination of the observed deviations has nevertheless triggered an intense activity on the phenomenological side in studies aiming at constraining the Wilson coefficients and probing possible contributions of New Physics \cite{flavio,straub,alguero}. The analysis of the late Run 2 data and the coming Run 3 data taking at LHC will be crucial to establish or infirm the observed anomalies. Furthermore, as those anomalies involve lepton flavour, a strengthening of the connection with neutrino physics is desirable \cite{boucenna,volkas}.

\end{document}